\documentclass[12pt]{article}

\pdfoutput=1

\usepackage{mathtools}
\usepackage{color}
\usepackage{amsmath}
\usepackage{graphicx}
\usepackage{amssymb}
\usepackage{amsthm}
\usepackage{upgreek}

\numberwithin{equation}{section}

\title{Spontaneously stochastic solutions \\ in one-dimensional inviscid systems}

\author{Alexei A. Mailybaev\footnote{Instituto 
Nacional de Matem\'atica Pura e Aplicada -- IMPA, 
Est. Dona Castorina 110, 22460-320 Rio de Janeiro, RJ, Brazil. 
E-mail: alexei@impa.br.} 
}

\date{}

\begin{document}

\maketitle

\begin{abstract}
In this paper, we study the inviscid limit of the Sabra shell model of turbulence, which is considered as a particular case of a viscous conservation law in one space dimension with a nonlocal quadratic flux function. We present a theoretical argument (with a detailed numerical confirmation) showing that a classical deterministic solution before a finite-time blowup, $t < t_b$, must be continued as a stochastic process after the blowup, $t > t_b$, representing a unique physically relevant description in the inviscid limit. This theory is based on the dynamical system formulation written for the logarithmic time $\uptau = \log(t-t_b)$, which features a stable traveling wave solution for the inviscid Burgers equation, but a stochastic traveling wave for the Sabra model. The latter describes a universal onset of stochasticity immediately after the blowup. 
\end{abstract}


\section{Introduction}

In this paper, we study the inviscid limit ($\nu \to 0^+$) for one-dimensional conservation laws of the form
\begin{equation}
	\frac{\partial u}{\partial t}+\frac{\partial f}{\partial x} = \nu \frac{\partial^2u}{\partial x^2},\quad 
	x,t \in \mathbb{R},
	\label{eqB0}
	\end{equation}
where $\nu \ge 0$ is the viscosity and the flux function $f$ is quadratic and nonlocal, i.e., $f = \iint K(y-x,z-x)u(y,t)u(z,t)dydz$. Such equations can be used as hydrodynamic models of turbulence, where the nonlocality of $f$ mimics the nonlocality of the pressure term in inviscid flows~\cite{frisch1995turbulence}. In fact, some of popular shell models of turbulence, which attracted a lot of interest due to their non-trivial behavior analogous to the developed hydrodynamic turbulence~\cite{biferale2003shell}, are strictly equivalent to Eq.~(\ref{eqB0}), see~\cite{mailybaev2014continuous}. In particular, this refers to the Sabra model of turbulence~\cite{l1998improved} studied in this paper. 

When $f = u^2/2$, Eq.~(\ref{eqB0}) represents the Burgers equation and its solution is well known. Inviscid solutions blow up in finite time forming a shock wave. A discontinuous (weak) solution at larger times is well-defined in the inviscid limit, see e.g.~\cite{dafermos2010hyperbolic}. When the flux function is nonlocal, a finite-time blowup in the inviscid system can be described using renormalization techniques~\cite{eggers2009role,mailybaev2012}. Our aim in this work is to demonstrate and explain the striking phenomenon, when a deterministic (classical) inviscid solution before the blowup continues spontaneously as a stochastic process for times after the blowup. 

Understanding of the stochasticity phenomenon proposed in this work is based on a combination of the two concepts: non-uniqueness and chaos. It is known that Lagrangian trajectories of a rough deterministic velocity field are non-unique~\cite{bernard1998slow,eijnden2000generalized,falkovich2001particles,kupiainen2003nondeterministic,eyink2015spontaneous}. 
The origin of this stochasticity is a violation of the Lipschitz condition, which ensures the uniqueness of solutions for differential equations, see e.g.~\cite{arnoldode}. In our system, the roughness necessary for such non-uniqueness is provided by the blowup phenomenon. 

It is widely accepted~\cite{frisch1995turbulence} that the developed turbulence is not just a finite-dimensional chaos phenomenon, due to a large (infinite as $\nu \to 0^+$) separation of scales both in space and time. These arguments are equally applied to the Sabra model of turbulence and the corresponding Eq.~(\ref{eqB0}).  We show, however, that the dynamical system approach can be used immediately after the blowup time $t_b$, if formulated for the logarithmic time $\uptau = \log(t-t_b)$. A crucial observation leading to the stochastic description is that the solution at every time $t > t_b$ undergoes an infinitely long chaotic evolution with respect to $\uptau$. We argue that this leads to the unique physically relevant description of the inviscid flow as a probability distribution for solutions $u(x,t)$ at $t > t_b$.

The paper is organized as follows. In Section~\ref{secBurgers} we show how the dynamics before and after the blowup in the inviscid Burgers equation can be translated into traveling wave solutions of respective renormalized systems. This representation is used in Section~\ref{sec3} to explain qualitatively the origin of the spontaneous stochasticity phenomenon. Section~\ref{sec2} introduces the Sabra model of turbulence and its continuous representation (\ref{eqB0}). Section~\ref{sec5} explains the universal self-similar structure of a finite-time blowup. Section~\ref{sec6} describes the solution at blowup time.  Section~\ref{sec7} demonstrates the universal emergence of a stochastic process from a deterministic blowup state. We end with the Conclusions. 

\section{Internal ``clock'' of the blowup}
\label{secBurgers}

\begin{figure}[h]
\centering
\includegraphics[width = 0.7\textwidth]{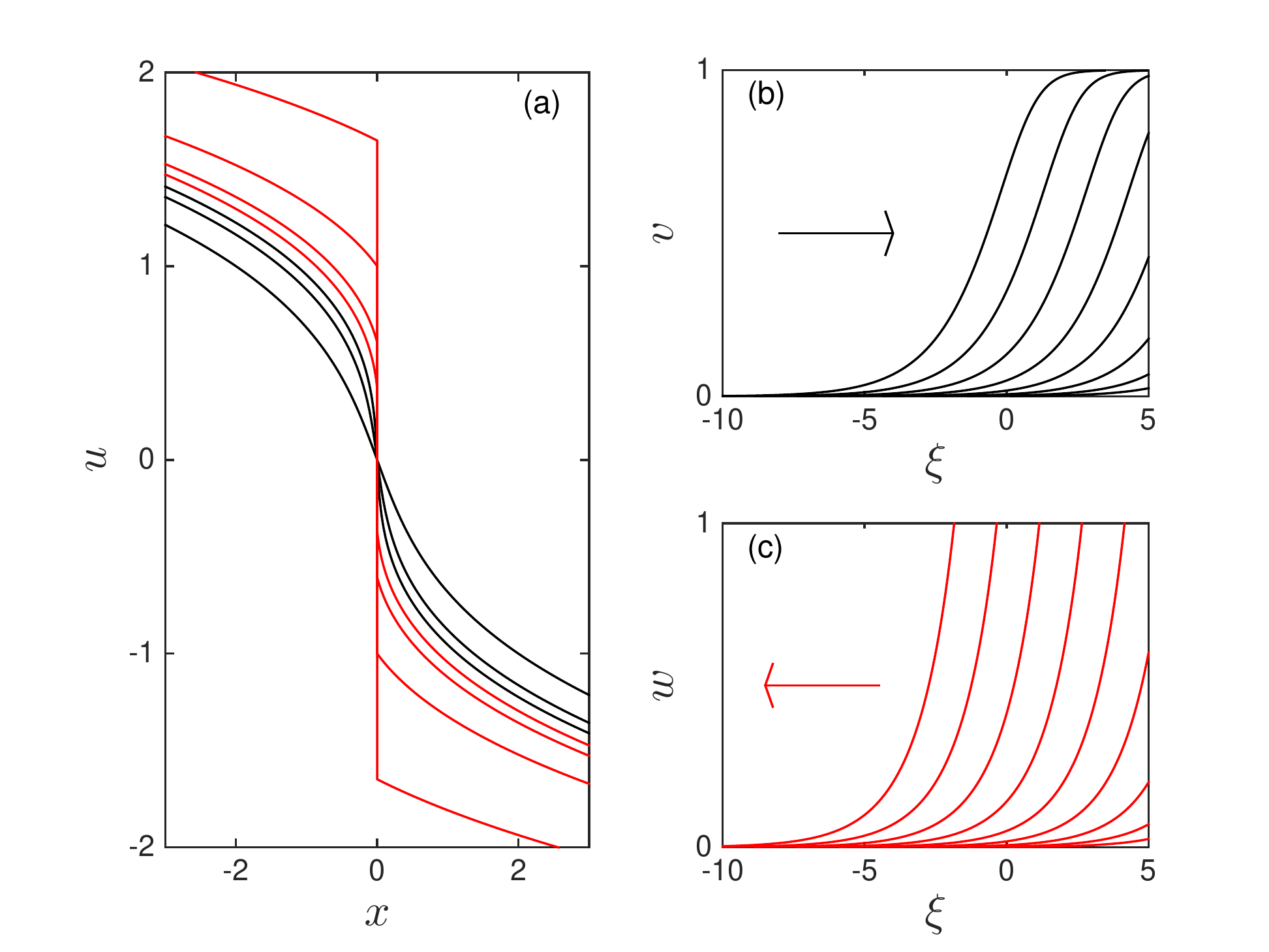}
\caption{Blowup in the inviscid Burgers equation with the initial condition $u_0(x) = -x+x^3$ at $t_0 = -1$. (a) Black curves show classical solutions at $t = -e^{-\tau}$ ($\tau = 0,1,2$) before the blowup. Red curves show shock wave solutions at  $t = e^{\uptau}$ ($\uptau = -2,-1,0,1$) after the blowup. Renormalized solutions: (b) $v(\xi,\tau)$ before the blowup ($\tau = 0,1,2,\ldots$) and (c) $w(\xi,\uptau)$ after the blowup ($\uptau = \ldots,-2,-1,0,1$) represent traveling waves moving with a constant speed in opposite directions.}
\label{fig1}
\end{figure}

In this section we describe how a finite-time blowup problem can be mapped into a problem of large-time behavior for a dynamical system. For this purpose, let us consider the Burgers equation (\ref{eqB0}), where the flux function $f = u^2/2$.
In the inviscid case ($\nu = 0$), a well-known classical (smooth) solution is given implicitly by 
	\begin{equation}
	u = u_0(x_0),\quad x = x_0+(t-t_0)u,
	\label{eqB1}
	\end{equation}
where $u(x,t_0) = u_0(x)$ is an initial condition and $x_0$ is an auxiliary variable.
Let us consider a solution, which blows up at finite time $t = t_b$. One can use a symmetry group of the Burgers equation, which includes shifts of origin, scale changes and the Galilean transformation, to simplify the blowup description. In generic case, this reduces the initial condition to the form $u_0(x) = -x+x^3+o(x^3)$ with $t_0 = -1$, see e.g.~\cite{pomeau2008wave,mailybaev2012}. Substituting this expression into Eq.~(\ref{eqB1}) and solving with respect to $x$ yields
	\begin{equation}
	x = ut-u^3+o(u^3).
	\label{eqB2}
	\end{equation}
The corresponding solution $u(x,t)$ blows up at $t_b = 0$, when $u(x,0) \approx -x^{1/3}$ has an infinite derivative at the origin, Fig.~\ref{fig1}(a). 

To simplify our further arguments, we ignore the $o(u^3)$ term in Eq.~(\ref{eqB2}). Then, for $x > 0$, we write this expression as
	\begin{equation}
	e^{-\xi} = ve^{-\xi}+v^3e^{3\tau-3\xi},
	\label{eqB3}
	\end{equation}
 where the renormalized time $\tau$, space variable $\xi$ and state $v$ are introduced as
 	\begin{equation}
	t = -e^{-\tau}, \quad
	x = e^{-\xi}, \quad
	u = -ve^{\tau-\xi}. 
	\label{eqB4}
	\end{equation}
Equation (\ref{eqB3}) yields
 	\begin{equation}
	v = F(\xi-a\tau),\quad a = 3/2,
	\label{eqB5}
	\end{equation}
where the function $F(\eta)$ is defined implicitly by the equation 
 	\begin{equation}
	1 = F+e^{-2\eta}F^3.
	\label{eqB5n}
	\end{equation}
We see that the blowup formation can be seen as a traveling wave (\ref{eqB5}) moving with the constant speed $a$ in the logarithmic space coordinate $\xi$ and time $\tau$, Fig.~\ref{fig1}(b). In this description, the limit $\tau \to \infty$ corresponds to the blowup time $t = -e^{-\tau}
 \to 0^-$, and the limit $\xi \to \infty$ yields the blowup location $x = e^{-\xi} \to 0$. Therefore, as the wave propagates to larger $\xi$, smaller values of $x$ are affected. At infinite time $\tau$, all scales get excited forcing the solution to blow up.
 
Using coordinates (\ref{eqB4}), we write
 	\begin{equation}
	\frac{\partial u}{\partial t} = -\left(\frac{\partial v}{\partial \tau}+v\right)e^{2\tau-\xi},\quad
	\frac{\partial u}{\partial x} = \left(\frac{\partial v}{\partial \xi}-v\right)e^{\tau}.
	\label{eqB5t}
	\end{equation}
Then the inviscid ($\nu = 0$) Burgers equation~(\ref{eqB0}) gets the form
 	\begin{equation}
	\frac{\partial v}{\partial \tau} = -v+v^2-v\frac{\partial v}{\partial \xi}.
	\label{eqB5b}
	\end{equation}
The traveling wave (\ref{eqB5}) is a solution of this equation, as it follows from the derivation and can also be checked directly.
A general blowup description should take into account the $o(u^3)$ correction in Eq.~(\ref{eqB2}). This yields a similar picture, but now the solution $v(\xi,\tau)$ becomes the traveling wave (\ref{eqB5}) asymptotically for large $\tau$, i.e., the blowup is associated with a stable solution $v = F(\xi-a\tau)$ of Eq.~(\ref{eqB5b}). Note that stability of this solution allows irrelevant unstable modes associated with the action of the symmetry group~\cite{eggers2009role}.  

For times after the blowup, $t > 0$, only a weak discontinuous solution exists. It can be uniquely defined as a solution of the viscous Burgers equation in the inviscid limit, $\nu \to 0^+$~\cite{dafermos2010hyperbolic}. This solution is given by the same Eq.~(\ref{eqB2}), but now one should choose positive (negative) values of $u$ before (after) a discontinuity $x = x_s(t)$, which propagates with the speed $dx_s/dt = \left[u(x_s^+)+u(x_s^-)\right]/2$, Fig.~\ref{fig1}(a). The discontinuity is located at the origin, $x_s(t) \equiv 0$, if one omits the $o(u^3)$ term in Eq.~(\ref{eqB2}).

The renormalized description for positive times is given by the variables
 	\begin{equation}
	t = e^{\uptau}, \quad
	x = e^{-\xi}, \quad
	u = -we^{-\uptau-\xi}, 
	\label{eqB6}
	\end{equation}
where we modified the first and the last expression as compared to Eq.~(\ref{eqB4}).
Analogous derivations yield the inviscid Burgers equation written in new coordinates as 
 	\begin{equation}
	\frac{\partial w}{\partial \uptau} = w+w^2-w\frac{\partial w}{\partial \xi},
	\label{eqB7b}
	\end{equation}
and Eq.~(\ref{eqB2}) provides the stable traveling wave solution
 	\begin{equation}
	w = G(\xi+\alpha\uptau),\quad \alpha = 3/2,
	\label{eqB7}
	\end{equation}
where $G(\eta)$ is a function defined by a positive root of the equation 
 	\begin{equation}
	1 = -G+e^{-2\eta}G^3.
	\label{eqB5m}
	\end{equation}
Now the blowup time, $t \to 0^+$, corresponds to $\uptau \to -\infty$. Hence, the increasing $\uptau$ describes the evolution from the blowup on.
The wave (\ref{eqB7}) travels with the speed $\alpha$ in the negative direction, from large positive $\xi$ (corresponding to small scales $x$), Fig.~\ref{fig1}(c). It describes a universal way how a discontinuity develops in a weak solution $u(x,t)$ immediately after the blowup. 

The presented description turns the evolution on each side of the blowup into a dynamical system, where a stable fixed-point (traveling wave) solution describes a universal shape of the blowup. The crucial element of this description is the logarithmic time $\tau$ used to unfold the finite-time singular behavior. This is not just an algebraic construction, but it reflects the infinite-dimensional nature of the blowup phenomenon. An every constant interval $\Delta\tau = (\log 2)/a$, in which the wave travels for a distance $\Delta\xi = \log 2$,  describes the time required for an excitation to evolve from the scale $x = e^{-\xi}$ to the twice smaller scale $x/2 = e^{-(\xi+\Delta\xi)}$. Thus, the time $\tau$ is merely measures the ``local'' characteristic time of the disturbance in its way from large to infinitely small scales at the blowup. Analogously, after the blowup, the time $\uptau$ is measured by the internal ``clock'' of the shock wave in its development from a point at the blowup to a finite size.
 
\section{Dynamical system description of blowup}
\label{sec3}

The transformation proposed in the previous section does not bring much new understanding for the Burgers equation, since the analytic solution is available, but it helps understanding the origin of the spontaneous stochasticity phenomenon studied below in this paper. A traveling wave solution is the simplest form of the large-time behavior for a translation-invariant autonomous dynamical system such as Eq.~(\ref{eqB5b}) or (\ref{eqB7b}). One can ask a question, what will happen if this solution gets unstable giving rise to a periodic or even chaotic attractor? For solutions before the blowup this is indeed possible if the nonlinear term $f$ in Eq.~(\ref{eqB0}) is nonlocal, as in continuous representations of shell models~\cite{mailybaev2014continuous}. This problem was studied in~\cite{mailybaev2012c} demonstrating different blowup scenarios corresponding to periodic, quasi-periodic and chaotic waves. These waves define an asymptotic form of a classical inviscid solution as it approaches the blowup, since large $\tau = -\log(t_b-t)$ correspond to $t \to t_b^-$. 

A very different situation is expected for a solution (\ref{eqB7}), which describes the unfolding of a blowup. This solution starts at $\uptau = \log(t-t_b) = -\infty$ corresponding to $t = t_b$. Therefore, an infinite interval (in terms of $\uptau$) preceeds any finite time after the blowup.  An example of the equation, where the attractor is a periodic wave was given in~\cite{mailybaev2015}. This means that there is a stable solution $w = G(\xi+\alpha\uptau,\uptau)$ such that $G(\eta,\uptau) = G(\eta,\uptau+\uptau_1)$ for some period $\uptau_1 > 0$ and any $\eta$ and $\uptau$. This solution represents a periodically pulsating wave traveling with an average speed $\alpha$ from large to small values of $\xi$. In fact, there is a family of solutions $w = G(\xi+\alpha\uptau+\xi_0,\uptau)$ defined up to a constant shift $\xi_0$, because the governing equation is translation invariant, see Eq.~(\ref{eqB7b}), for example.  It was shown that a specific value of $\xi_0$ is chosen if one defines a solution in the inviscid limit $\nu_n \to 0^+$ for a specific sequence of viscosities. Any value of $\xi_0$ can be obtained in this way, leading to the non-uniqueness (an infinite number) of \textit{physically relevant} inviscid solutions. 

\begin{figure}
\centering
\includegraphics[width = 0.9\textwidth]{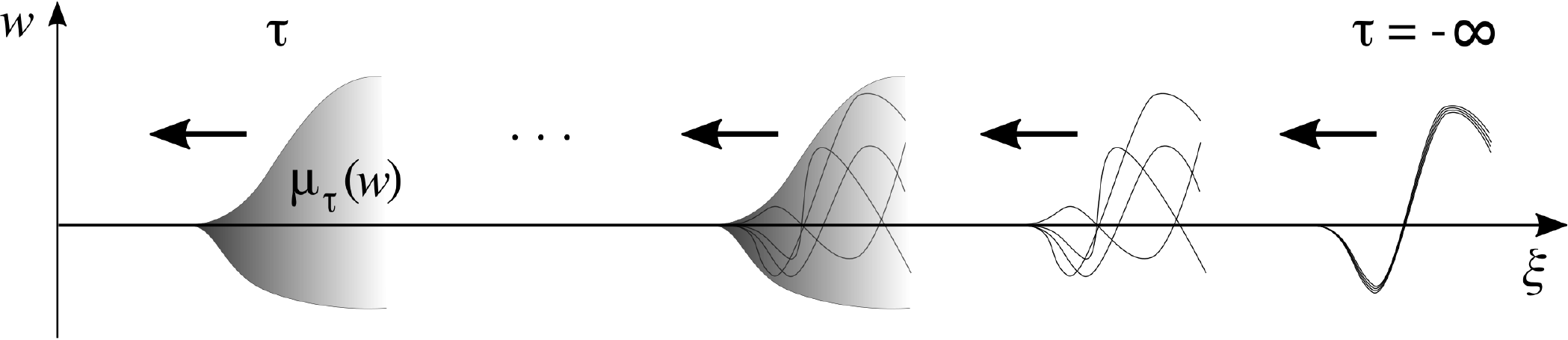}
\caption{Schematic picture of the formation of a traveling-wave probability measure $\mu_{\uptau}(w)$ for renormalized chaotic solutions $w(\xi,\uptau)$ from a blowup state at $\uptau = -\infty$.}
\label{fig2}
\end{figure}

In this paper we show that the Sabra shell model of turbulence~\cite{l1998improved}, which is equivalent to system (\ref{eqB0}) with a nonlocal quadratic flux function given below by Eqs.~(\ref{eq8}) and (\ref{eq8b}),  provides an example in which the blowup unfolding is given by a chaotic wave. This means that $w = G(\xi+\alpha\uptau,\uptau)$, where $G(\eta,\uptau)$ is characterized for large $\uptau$ by a chaotic attractor. The blowup imposes a specific initial condition for the inviscid solution $w$ at $\uptau \to -\infty$. But, due to exponential divergence of trajectories with close initial conditions (the famous butterfly effect), we cannot choose any particular  solution at finite $\uptau$, see Fig.~\ref{fig2}. Namely, even if a particular inviscid solution is chosen by some viscous regularization procedure, with a discrete subsequence of viscosities $\nu_n \to 0^+$~\cite{lions1996mathematical}, an arbitrarily small perturbation will provide a totally different solution. 
Thus, no physically relevant deterministic solution can be expected in a vanishing viscosity limit. The limiting object is a chaotic attractor with an invariant measure, which describes probability for different observable solutions, Fig.~\ref{fig2}. 

We are led to the surprising conclusion: the inviscid solution becomes stochastic at every time after the blowup despite the governing equation (\ref{eqB0}) looks fully deterministic! This is not just the non-uniqueness phenomenon for weak solutions. On the contrary, one can expect a unique solution as a probability distribution, because this distribution is associated with the invariant measure of a chaotic attractor. 

Our argument above suggests, and we will provide a detailed numerical evidence, that a physically relevant inviscid solution at times after the blowup should be defined as a measure $d\mu_{\uptau}(w)$, which describes a probability distribution of solutions $w(\xi,\uptau)$ at fixed $\uptau = \log (t-t_b)$. The measure $d\mu_{\uptau}(w)$ has the form of a traveling wave moving with a constant speed $\alpha$ in logarithmic coordinates $(\xi,\uptau)$. This traveling wave connects a deterministic blowup state on one side (deterministic ``past'' of the solution) and a stochastic state on the other side. A steady motion of such a wave in the direction of smaller $\xi$ describes the propagation of stochasticity from arbitrarily small to finite scales $x = e^{-\xi}$, Fig.~\ref{fig2}. 

An important theoretical implication is the revision of a viscous regularization procedure. A small viscosity present in a physical system suppresses the dynamics at sufficiently large $\xi$ in Fig.~\ref{fig2}. This yields a deterministic system with a unique solution globally in time (though this is not proved for the Navier-Stokes equations). Thus, a formal way that leads to a stochastic limit must include a random component in a vanishing regularization term. This may be a small-scale noise~\cite{lorenz1969predictability,leith1972predictability,ruelle1979microscopic,eyink1996turbulence} or a random viscosity perturbation~\cite{mailybaev2015b}, which vanish together with the viscosity, as discussed earlier in the context of the inverse cascade of stochasticity in the developed turbulence.

\section{Sabra shell model of turbulence and its continuous representation}
\label{sec2}

Shell models are obtained by reducing the fluid dynamics equations (e.g., Naver--Stokes equations) to a discrete sequence of shells $|\mathbf{k}| = k_n$ in the Fourier space for the geometric 
progression of wavenumbers $k_n = k_0\lambda^n$, $n = 1,2,3,\ldots$, where $\lambda > 1$ is the inter-shell ratio. Relating the spatial scale to the wavenumber, $\ell \sim k_n^{-1}$, one associates large scales with the initial shells, $n \sim 1$, while small scales are given by large $n$. The ``flow'' is then described by complex  variables $u_n(t)$, called shell speeds. Though no quantitative relation of such models to the original flow equations is expected, the shell models possess a number of non-trivial (not yet fully understood) properties of turbulence like, e.g., the intermittency phenomenon. Clear advantages of shell models are their relative simplicity and convenience for accurate numerical simulations.

In this paper, we consider the Sabra shell model~\cite{l1998improved}, which was obtained after some improvements of the Gledzer--Ohkitani--Yamada (GOY) model~\cite{gledzer1973system,ohkitani1989temporal}. This model is given by an infinite system of equations 
	\begin{equation}
	\frac{d u_n}{dt} = N_n[u]-\nu k_n^2u_n, \quad n = 1,2,3,\ldots,
	\label{eq1}
	\end{equation}
where $\nu \ge 0$ is a viscosity parameter and the quadratic nonlinear term is defined as
	\begin{equation}
		N_n[u] = i\left(k_{n+1}u_{n+2}u_{n+1}^*-\frac{1}{2}k_nu_{n+1}u_{n-1}^*
		+\frac{1}{2}k_{n-1}u_{n-1}u_{n-2}\right), 
	\label{eq4}
	\end{equation}
with the stars denoting the complex conjugation.
The model must be suppled with large-scale boundary conditions for the shell speeds $u_{0}$ and $u_{-1}$.
One typically adds forcing terms $f_n$ acting at large scales (small $n$) in Eq.~(\ref{eq1}). For simplicity, but with no conceptual difference for the results, we will not consider such forcing; an external excitation can be produced by the boundary conditions as it is typical in fluid dynamics. The model possesses two inviscid invariants, the energy $E = \sum |u_n|^2$ and the helicity $H = \sum (-1)^nk_n|u_n|^2$ (the summation over all $n$ is assumed).

Solutions of viscous shell models exist and unique globally in time~\cite{constantin2006analytic}. 
For the inviscid models, i.e., with $\nu = 0$ in Eq.~(\ref{eq1}), the criterion of existence and uniqueness of the solution requires a finite (enstrophy) norm 
	\begin{equation}
	\omega = \left(\sum k_n^2|u_n|^2\right)^{1/2}.
	\label{eq5}
	\end{equation}
If $\omega = \infty$, the solution can be defined in a weak sense, but its uniqueness is not known~\cite{constantin2007regularity}. 

We will also consider an entirely different derivation of the Sabra model~\cite{mailybaev2014continuous}. For the specific value of the inter-shell ratio, 
\begin{equation}
\lambda = \sqrt{2+\sqrt{5}} \approx 2.058, 
\label{eq8g}
\end{equation} 
the Sabra model can be derived rigorously from the one-dimensional viscous conservation law (\ref{eqB0}). Here the nonlocal flux function is given by
\begin{equation}
f = \iint K(y-x,z-x)u(y,t)u(z,t)dydz
\label{eq8}
\end{equation} 
with the kernel 
\begin{equation}
K(y,z) 
= \frac{K_\psi(y,z)+K_\psi(z,y)}{4\pi},\quad
K_\psi(y,z) 
= \frac{2\sigma}{(\sigma y-z)^2}-\frac{\sigma^2}{(\sigma^2y-z)^2}
+\frac{\sigma}{(\sigma y+z)^2},
\label{eq8b}
\end{equation} 
where $\sigma = (1+\sqrt{5})/2$ is the golden ratio.
Singular integrals in Eq.~(\ref{eq8}) must be taken with the Hadamard regularization.
Let $u(x,t)$ be a solution of Eqs.~(\ref{eqB0}), (\ref{eq8}), (\ref{eq8b}), and $\hat{u}(k,t) = \int u(x,t)e^{-ikx}dx$  its Fourier transform. Then, for every fixed $1 \le k_0 < \lambda$, the functions
\begin{equation}
u_n(t) = k_n^{1/3}\hat{u}\left(k_n^{2/3},t\right),\quad k_n = k_0\lambda^n, 
\label{eqS1}
\end{equation} 
yield a solution of the Sabra model (\ref{eq1}), (\ref{eq4}), see~\cite{mailybaev2014continuous}. Therefore, Eq.~(\ref{eqB0}) in the Fourier representation splits into a family of independent Sabra models parametrized by $k_0$. 

Note that the nonlocal quadratic term (\ref{eq8}) is natural for a model of turbulence, because it reflects a nonlocal character of the pressure term in incompressible flows. In the next sections, we study the Sabra model, taking into account that the conclusions are automatically valid for its one-dimensional continuous representation (\ref{eqB0}). 

\section{Self-similar dynamics before blowup}
\label{sec5}

Let us consider the inviscid Sabra model
\begin{equation}
\frac{du_n}{dt} = N_n[u], \quad n = 1,2,3,\ldots.
\label{eq10}
\end{equation}
For a generic initial condition, with a finite norm $\omega_0 = \omega(0)$, the inviscid solution of the Sabra model blows up in finite time, i.e., $\omega(t) \to \infty$ as the solution approaches the blowup time $t \to t_b^-$. The local analysis presented in this section follows the construction of~\cite{dombre1998intermittency}, where the reader can see the derivations in more detail; see also further developments in~\cite{mailybaev2012,mailybaev2012c,mailybaev2013blowup}. For this purpose, we introduce the rescaled shell variables $v_n$ and time $\tau$ as
\begin{equation}
v_n = \frac{ik_nu_n}{\omega/\omega_0},\quad 
\frac{d\tau}{dt} = \frac{\omega}{\omega_0}.
\label{eq11}
\end{equation}
According to the definition (\ref{eq5}), new variables conserve the sum $\sum |v_n|^2 = \omega_0^2 = \mathrm{const}$.
Such transformation is well defined at times before the blowup and leads to the following equations~\cite{dombre1998intermittency,mailybaev2013blowup}
\begin{equation}
\frac{d{v}_n}{d\tau} = P_n[v]-Av_n, \quad n = 1,2,3,\ldots,
\label{eq13}
\end{equation}
where
\begin{equation}
P_n[v] =  -\frac{1}{\lambda^2}v_{n+2}v_{n+1}^*+\frac{1}{2}v_{n+1}v_{n-1}^*
		+\frac{\lambda^2}{2}v_{n-1}v_{n-2}
\label{eq12}
\end{equation}
and $A = d \log\omega/d\tau$.
This expression for $A$ can be written using Eqs.~(\ref{eq5}) and (\ref{eq10})--(\ref{eq12}) as
\begin{equation}
A = \frac{\mathrm{Re}\sum v_n^*P_n[v]}{\sum |v_n|^2}.
\label{eq12x}
\end{equation}

\begin{figure}
\centering
\includegraphics[width = 1\textwidth]{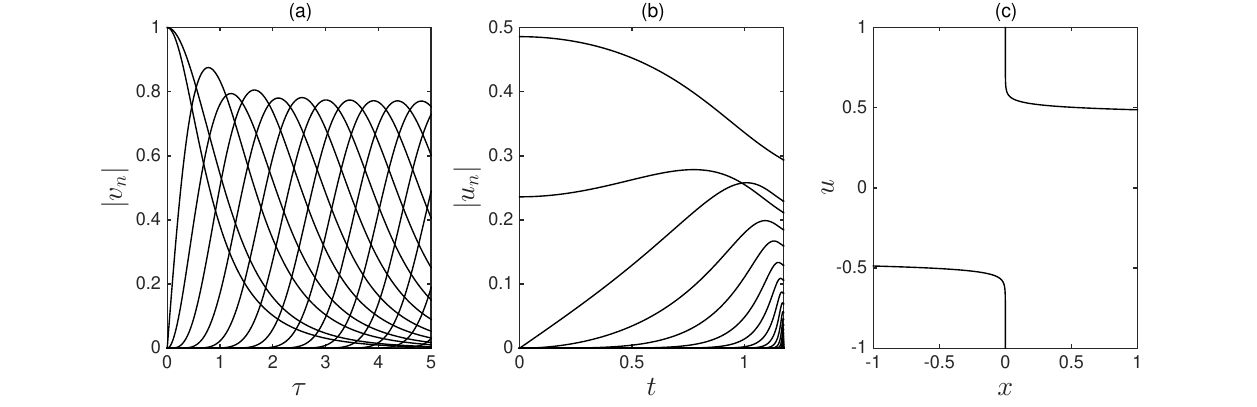}
\caption{(a) Renormalized variables $v_n(\tau)$ forming a traveling wave for large $\tau$; the shell number $n = 1,2,\ldots$ increases from the left to the right. The initial conditions are $v_1 = v_2 = 1$ (with zeros for other variables) at $\tau = 0$. (b) Corresponding dynamics of the original variables $u_n(t)$ developing into a self-similar blowup. (c) Physical space representation $u(x,t)$ of the solution $u_n(t)$ at blowup time.}
\label{fig3}
\end{figure}

Equation (\ref{eq13}) is translation-invariant with respect to the shell number $n$, i.e., it does not change under the transformation $v_n \mapsto v_{n+j}$ for any $j$ (except in the region near the boundary condition). It was shown numerically~\cite{mailybaev2013blowup}, that Eq.~(\ref{eq13}) of the Sabra model has a stable traveling wave solution, which can be written as
\begin{equation}
v_n(\tau) = e^{i\theta_n}V(n-a\tau),
\label{eq14}
\end{equation}
where $\theta_n$ are arbitrary phases (resulting from an action of the symmetry group) such that $\theta_n = \theta_{n-1}+\theta_{n-2}$. The wave (\ref{eq14}) propagates with the constant speed $a$ in the direction of large $n$ (small scales), Fig.~\ref{fig3}(a).

For the traveling wave solution, the function $A(\tau)$ given by Eq.~(\ref{eq12x}) is periodic with the period $\tau = 1/a$. Hence, using the relation $A = d \log\omega/d\tau$, one obtains
 \begin{equation}
\omega = \omega_0\exp\left(\int_0^{\tau}A(\tau')d\tau'\right) = \omega_0h(\tau)\lambda^{za\tau}, 
\quad
z = \frac{1}{\log \lambda}\int_0^{1/a}A(\tau)d\tau,
\label{eq16b}
\end{equation}
where $h(\tau)$ is a positive $(1/a)$-periodic function. The scaling exponent $z$ is determined by the stable traveling wave solution and, hence, it is universal (independent of initial conditions). Numerical simulations in the case (\ref{eq8g}) yield $z \approx 0.6975$. 
As long as $z > 0$, the second expression in (\ref{eq11}) ensures that the original time has a finite limit, $t \to t_b < \infty$, as $\tau \to \infty$. At the same time, the value $\omega \sim \lambda^{za\tau} \to \infty$, i.e., the system blows up at finite time $t_b$.

In the original variables (\ref{eq11}), the asymptotic traveling wave solution (\ref{eq14}) yields~\cite{dombre1998intermittency,mailybaev2013blowup}
\begin{equation}
u_n(t) = -ie^{i\theta_n}k_n^{z-1}U(k_n^{z}(t-t_b)), \quad t < t_b,
\label{eq15}
\end{equation}
where 
\begin{equation}
U(t-t_b) = h(\tau)\lambda^{za\tau}V(-a\tau). 
\label{eq16}
\end{equation}
Thus, a traveling wave solution (\ref{eq14}) leads to the self-similar asymptotic behavior at times preceding the blowup as described by Eq.~(\ref{eq15}), see Fig.~\ref{fig3}(b). The function $U(t)$ has a finite limit $U(0)$ as $t \to -0$, which imposes the decay rate $V(\eta) \sim \lambda^{z\eta}$ as $\eta \to -\infty$~\cite{dombre1998intermittency}.

Using Eq.~(\ref{eq16b}) and $k_n = k_0\lambda^n$, we write the asymptotic scaling of variables (\ref{eq11}) as
\begin{equation}
u_n \propto \lambda^{za\tau-n}v_n,\quad
t-t_b \propto -\lambda^{-za\tau}.
\label{eq11b}
\end{equation}
Additionally, the physical scale of shell $n$ is expressed  as
\begin{equation}
\ell \propto k_n^{-1} \propto \lambda^{-n}.
\label{eq11c}
\end{equation}
The comparison of Eqs.~(\ref{eq11b}), (\ref{eq11c}) with Eq.~(\ref{eqB4}) shows that our description of the blowup in the Sabra model is analogous to the one for the Burgers equation in Section~\ref{secBurgers}. Here $n$ stands for the logarithmic space variable $\xi$. The blowup is associated with a wave traveling to large $n$ with a constant speed in the logarithmic time $\tau$. 

Note that the shapes of the two traveling waves in Figs.~\ref{fig1}(b) and~\ref{fig3}(a) look different, because the Burgers description was based on physical space representation, while the Sabra model corresponds to the Fourier-transformed equations. Considering the Burgers equation in Fourier space yields the description, which looks rather similar to the Sabra model~\cite{mailybaev2012}. 

\section{The blowup state}
\label{sec6}

Exactly at the blowup time, $t = t_b$, expression (\ref{eq15}) yields $u_n(t) = -ie^{i\theta_n}k_n^{z-1}U(0)$. With no loss of generality, we can drop the coefficients assuming that
\begin{equation}
u_n(t_b) = -ik_n^{z-1}, 
\label{eq22}
\end{equation}
which can be obtained by using a symmetry group of the Sabra model~\cite{mailybaev2013blowup}. Eq.~(\ref{eq22}) describes the asymptotic inviscid state for large shell numbers $n$. Using Eq.~(\ref{eqS1}), one recovers the function $\hat{u}(k,t_b) = -ik^{-\beta}$ for $k > 0$ and $\beta = 2-3z/2 \approx 0.954$, which is a Fourier transposed solution of the continuous representation for the Sabra model. As a Fourier transform of a real function, it extends to negative $k < 0$ as $\hat{u}(k) = \hat{u}^*(-k) = i|k|^{-\beta}$. The inverse Fourier transform yields~\cite{bateman1954tables}
\begin{equation}
u(x,t_b) = \frac{\Gamma(1-\beta)}{\pi}\cos\left(\frac{\beta\pi}{2}\right)|x|^{\beta-1}\mathrm{sgn}\,x. 
\label{eq22b}
\end{equation}
This function is shown in Fig.~\ref{fig3}(c) and it represents a discontinuity of the solution in physical space, which is created at the blowup time. As $\beta$ is close to $1$, function (\ref{eq22b}) is close to a discontinuity (shock), but has infinite limits at $x = 0$. One can also find numerically the physical space representation $u(x,t)$ of the asymptotic relation (\ref{eq15}), which describes a self-similar formation of a singularity (\ref{eq22b}) in a classical solution~\cite{mailybaev2014continuous}.      

\section{Spontaneously stochastic dynamics after blowup}
\label{sec7}

In order to study the behavior after blowup, we introduce the new variables
\begin{equation}
t = t_b+\lambda^{\uptau},\quad
u_n = -ik_n^{-1}\lambda^{-\uptau}w_n = -ik_0^{-1}\lambda^{-\uptau-n}w_n.
\label{eqT1}
\end{equation}
Together with the shell scale $\ell \propto k_n^{-1} = k_0^{-1}\lambda^{-n}$, expressions (\ref{eqT1}) follow the analogous definition (\ref{eqB6}) for the Burgers equation. For new variables, the inviscid Sabra model (\ref{eq10}), (\ref{eq4}) takes the form
	\begin{equation}
	\frac{d{w}_n}{d\uptau} = \left(w_n-\frac{1}{\lambda^2}w_{n+2}w_{n+1}^*+\frac{1}{2}w_{n+1}w_{n-1}^*
		+\frac{\lambda^2}{2}w_{n-1}w_{n-2}\right)\log\lambda. 
	\label{eqS2}
	\end{equation}
This equation is autonomous and translation-invariant, thus, it allows traveling-wave type of solutions. Condition (\ref{eq22}) at $t = t_b$ must be satisfied in the limit $\uptau \to -\infty$. For this limit, the second relation in Eq.~(\ref{eqT1}) with $k_n = k_0\lambda^n$ defines 
\begin{equation}
w_n = ik_n\lambda^{\uptau}u_n \to k_n^{z}\lambda^{\uptau} = k_0^{z}\lambda^{z\left(n+\alpha\uptau\right)},\quad
\alpha = 1/z,\quad \uptau \to -\infty.
\label{eqS3}
\end{equation}
This expression determines the speed $\alpha = 1/z \approx 1.4337$ of a traveling wave, with the direction of motion from larger to smaller shell numbers $n$.

 \begin{figure}
\centering
\includegraphics[width = 0.85\textwidth]{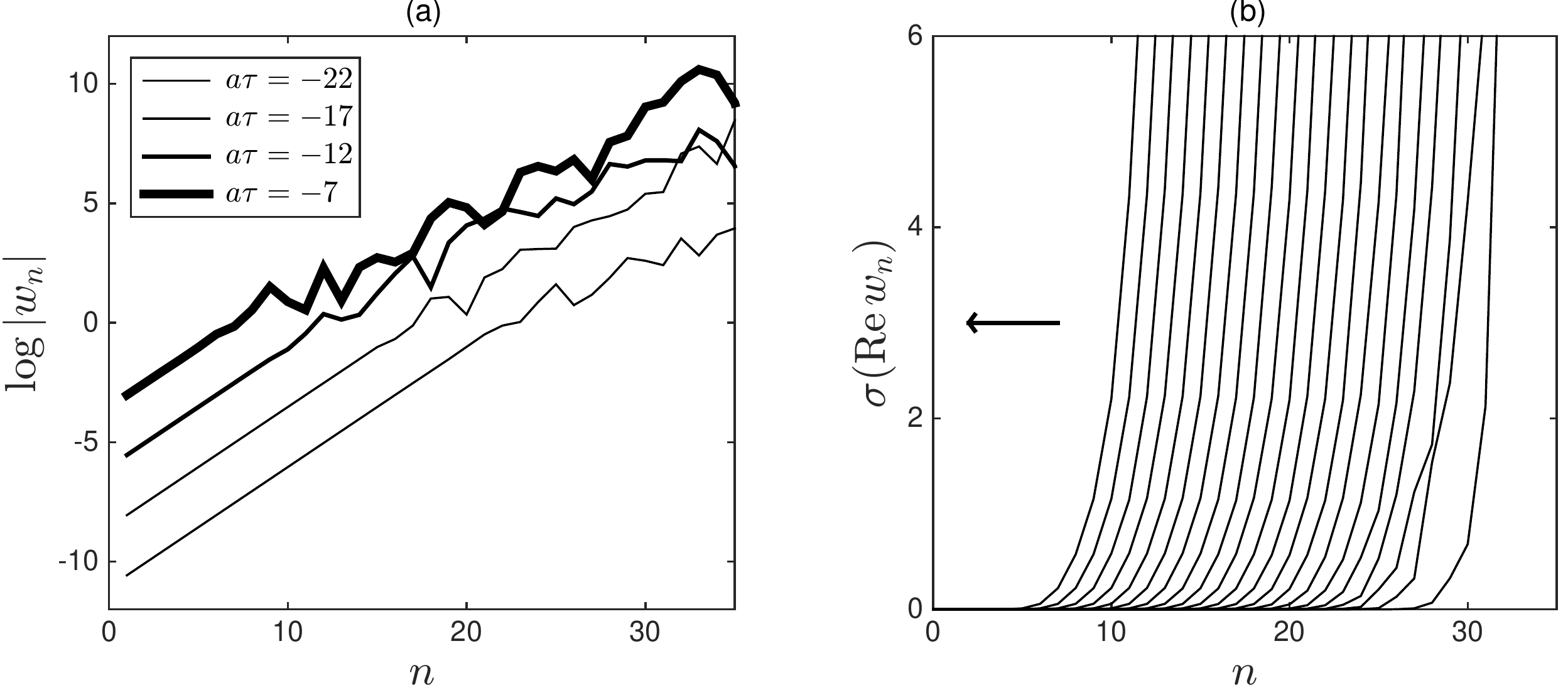}
\caption{(a) Chaotic dynamics of variables $w_n$ shown at different times $\uptau = \log_{\lambda} t$, where $\uptau = -\infty$ corresponds to blowup state (\ref{eq22}). The data is obtained from a single integration of the Sabra model with viscosity $\nu = 10^{-15}$. (b) Standard deviation of $\mathrm{Re}\,w_n$ at $\alpha\uptau = -27,-26,\ldots,-7$ (increasing time is indicated by an arrow). The graphs form a traveling wave moving in the direction of smaller $n$ with the constant speed $\alpha$.}
\label{fig4}
\end{figure}

Numerical simulations suggest that the dynamics described by Eq.~(\ref{eqS2}) (with viscous regularization) is chaotic, which is a well-known fact for the Sabra model, Fig.~\ref{fig4}(a). In this case, our argument in Section~\ref{sec3} suggests that the inviscid solution should be understood in the probabilistic sense, i.e., as a measure $d\mu_{\uptau}(w)$ describing a probability distribution for the infinite sequence $w = (w_1,w_2,\ldots)$ at given $\uptau$. This measure should be obtained in the inviscid limit, which includes a small-scale random perturbation. In order to verify this hypothesis, we found the statistical distribution numerically. We took the asymptotic blowup state~(\ref{eq22}) as the initial condition at $t = t_b = 0$. A very small viscosity is set to $\nu = 10^{-15}$. Also, a small perturbation is applied in the viscous range, $u_{36}(0) = (-i+0.01x)k_{36}^{z-1}$, with a random real number $x$ uniformly distributed in the interval $[-1,1]$ (a specific form and magnitude of this perturbation does not affect the results). Eqs.~(\ref{eq1}) and (\ref{eq4}), with $\lambda$ given by Eq.~(\ref{eq8g}) and the total number of shells $n = 45$, are integrated numerically with high accuracy. We performed $10^4$ simulations for different values of the random number $x$. These numerical simulations feature the viscous range for shells $n \sim 35$. Thus, we can observe the inviscid dynamics in the inertial interval of shells $n \lesssim 30$. 

\begin{figure}
\centering
\includegraphics[width = 0.99\textwidth]{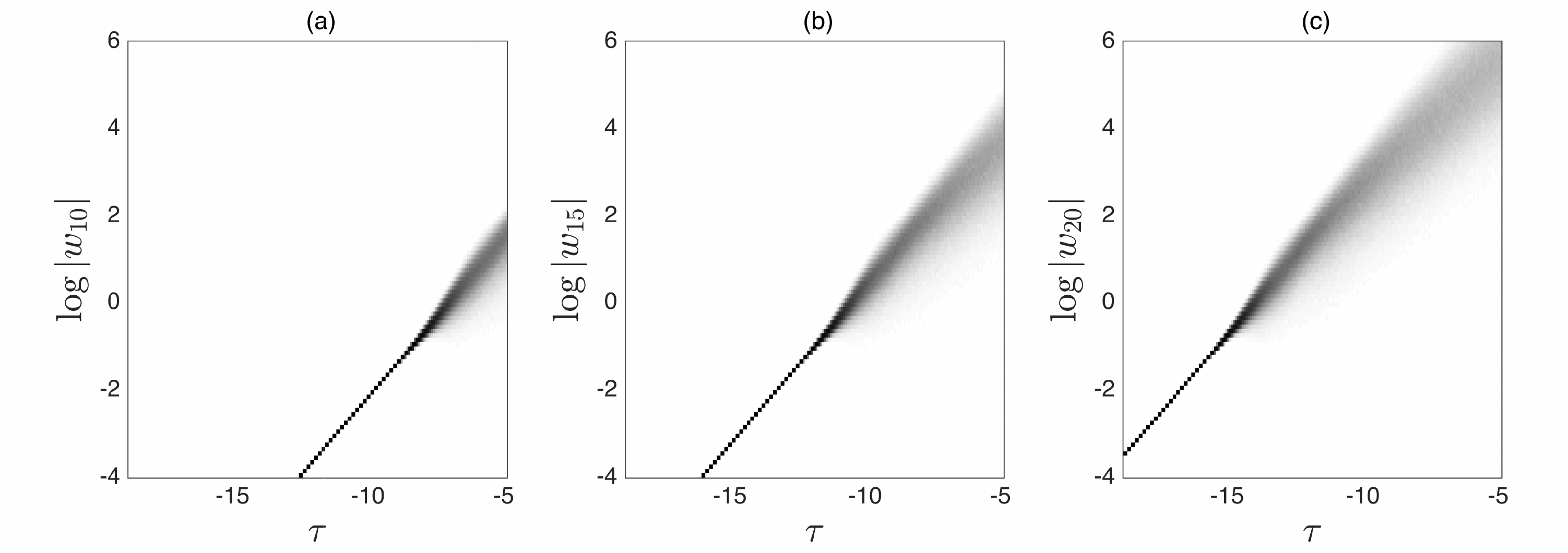}
\caption{PDFs of $\log |w_{n}|$ for (a) $n = 10$, (b) $n = 15$, (c) $n = 20$ as functions of $\uptau = \log_\lambda t$. Darker color means larger probability. These functions represent a traveling wave in $(n,\tau)$ coordinates, with a deterministic blowup state for $\uptau \to -\infty$.}
\label{fig5}
\end{figure}

Fig.~\ref{fig4}(b) shows the standard deviation of $\mathrm{Re}\,w_n$ with increasing $\uptau$. The figure demonstrates the formation of a stable traveling wave moving with the constant speed $\alpha$ in the direction of small $n$, i.e., from small to large scales. Note a similarity of Fig.~\ref{fig4}(b) with the analogous graph for the deterministic $w(\xi,\uptau)$ of the Burgers equation in Fig.~\ref{fig1}(c).  For the probability measure $\mu_{\uptau}(w)$, the traveling wave condition implies
\begin{equation}
\mu_{\uptau+\uptau_0}(w) = \mu_{\uptau}(Tw),
\label{eqS4}
\end{equation}
 where $\uptau_0 = 1/\alpha = z$ is the time period, in which the wave travels for a distance of one shell number, and $T:(w_1,w_2,\ldots) \mapsto (w_2,w_3,\ldots)$ is the corresponding translation operator. Fig.~\ref{fig5} shows probability density functions (PDFs) of $\log |w_n|$ for the shells $n = 10,\,15,\,20$ at different times $\uptau$, which are in full agreement with the traveling wave condition (\ref{eqS4}). For each shell, the stochastic component grows with $\uptau$ in the same way but with a shift in $\uptau$. The stochasticity of equal intensity is developed earlier at larger shell numbers (smaller scales) and later for smaller shell numbers (larger scales). The limit $\uptau \to -\infty$, corresponding to $t \to t_b^+$, describes a deterministic ``past'' of the solution given by Eq.~(\ref{eqS3}), and it is clearly seen as a straight sold line in Fig.~\ref{fig5}. 
 
\begin{figure}[t]
\centering
\includegraphics[width = 0.85\textwidth]{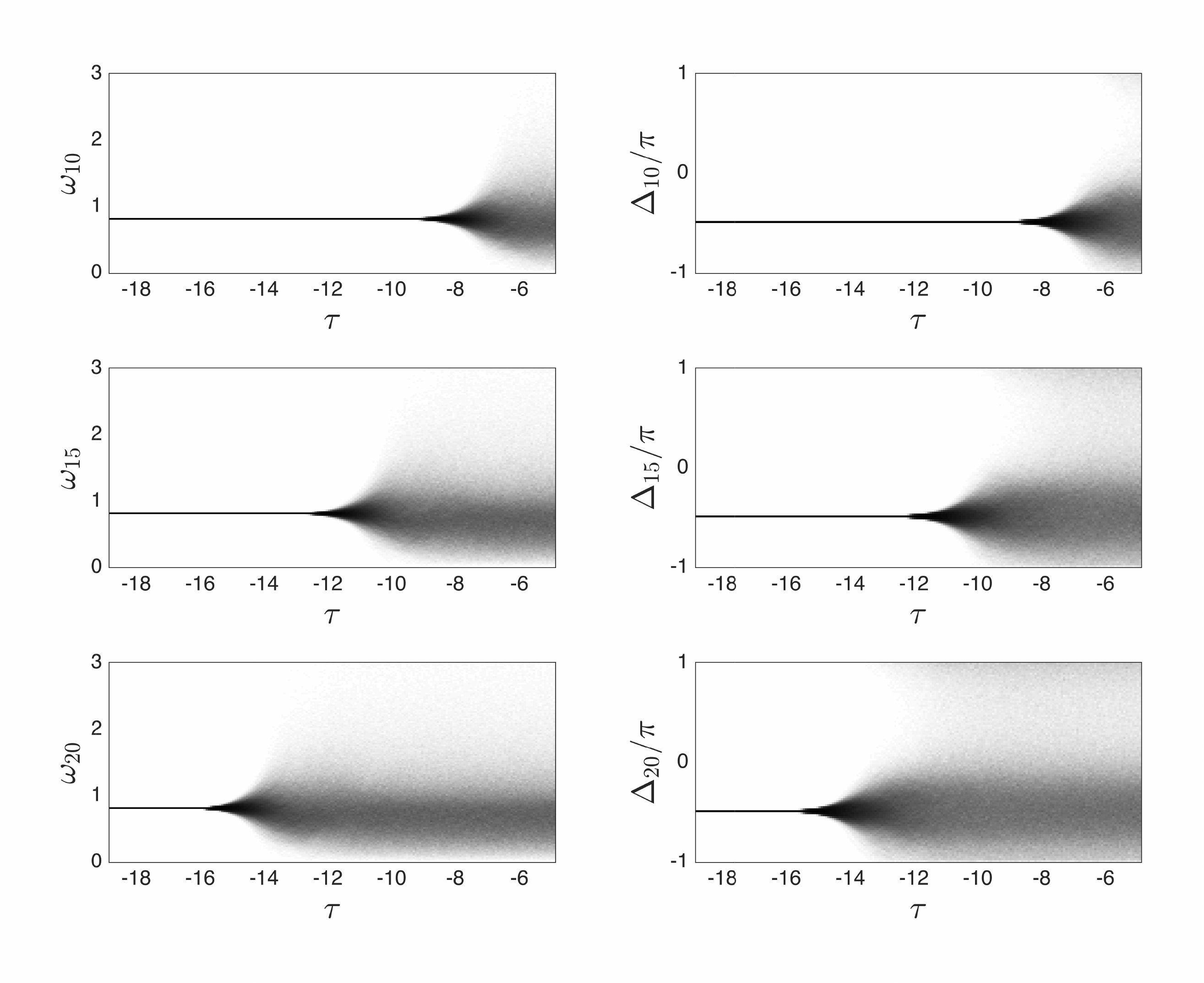}
\caption{PDFs of the multiplier $\omega_n$ (left column) and the phase $\Delta_n$ (right column) for the shells $n = 10,\,15,\,20$ as functions of $\uptau = \log_\lambda t$. Darker color means larger probability. These functions represent a traveling wave  with a constant deterministic state for small $\uptau$ and a constant stochastic state for large $\uptau$.}
\label{fig6}
\end{figure}

\begin{figure}
\centering
\includegraphics[width = 0.9\textwidth]{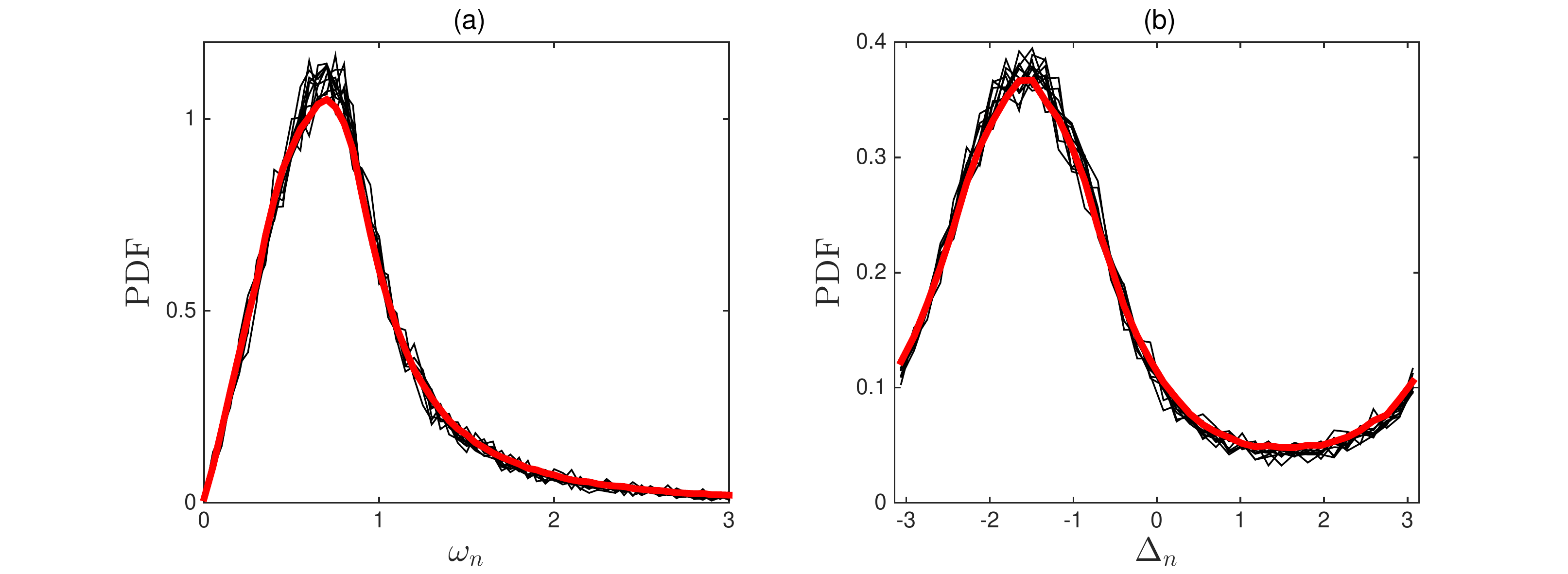}
\caption{PDFs of (a) the multiplier $\omega_n$ and (b) the phase $\Delta_n$. Thin black lines are obtained from the simulations for the shells $n = 15,\ldots,25$ at final time in Fig.~\ref{fig6}. All the curves collapse onto the PDF corresponding to the stationary turbulent regime (bold red curves)~\cite{eyink2003gibbsian}.}
\label{fig7}
\end{figure}

An interesting representation of the traveling probability measure can be obtained using the shell speed multipliers, which include the factors and phases defined as~\cite{benzi1993intermittency,eyink2003gibbsian}
	\begin{equation}
	\omega_n = |u_n/u_{n-1}|,\quad \Delta_n = \arg(u_{n-2}u_{n-1}u_n^*).
	\label{eqF1}
	\end{equation}
According to the Kolmogorov hypothesis~\cite{kolmogorov1962refinement,chen2003kolmogorov}, these variables have universal statistics for the stationary developed turbulence. Condition (\ref{eqS4}) implies that the PDFs of the random variables $(\omega_n,\Delta_n)$ have the form of a traveling wave. These PDFs are shown in Fig.~\ref{fig6}. They not only confirm the traveling wave form of the solution, but also demonstrate a stationary stochastic state on the right side, $\uptau \to \infty$. This state corresponds to the stationary developed turbulence, according to the Kolmogorov hypothesis, which we confirm in Fig.~\ref{fig7}. 

Finally, we performed additional tests, where we followed the solution from large-scale initial conditions, both before and after the blowup. In order to see the stochastic inviscid limit, we assumed a small perturbation of the viscosity $\nu = 10^{-15}(1+0.01x)$. The obtained results lead to the probability distribution in the form of a traveling wave with no noticeable difference (within the numerical accuracy) in comparison with Figs.~\ref{fig5} and \ref{fig6}, confirming the asymptotic universality of the stochastic solution.

\section{Conclusions}

In this work, we studied solutions for one-dimensional models of hydrodynamic type, $u_t+f_x = \nu u_{xx}$, where the quadratic flux function $f$ is nonlocal. Our main focus is the behavior of solutions in inviscid limit near the blowup time $t_b$. We showed that an asymptotic evolution before the blowup can be mapped to an autonomous dynamical system with the logarithmic temporal variable $\tau = -\log (t_b-t)$. Similarly, a dynamical system can be introduced after the blowup with the logarithmic time $\uptau = \log (t-t_b)$. For the Burgers equation, with $f = u^2/2$, these two dynamical systems have stable traveling wave solutions describing a universal form of shock formation. However, chaotic waves may appear for models with a nonlocal flux function $f$, as we demonstrated for the Sabra shell model of turbulence and its one-dimensional representation. This chaotic behavior triggers a spontaneous probabilistic description for the system solutions.

A crucial element of our analysis is the existence of a probability measure in the form of a traveling wave, which moves from small to large scales with constant profile and speed in the logarithmic space-time. This wave has the blowup state on one side, describing a deterministic past at $\uptau = -\infty$, and the developed turbulent state on the other side, describing a stochastic future at finite $\uptau$. The semi-infinite interval $(-\infty,\,\uptau]$ of the chaotic dynamics collapses into a finite interval $(t_b,\,t]$ for physical time. As a result, we are led to a unique physically relevant inviscid solution in the form of a probability distribution. This yields the main conclusion of this paper: a deterministic (classical) inviscid solution before the blowup continues spontaneously as a stochastic process for times after the blowup. Detailed numerical simulations for the Sabra model of turbulence demonstrate an excellent agreement with this theory.

The proposed probabilistic definition of the solution leads to important requirements, which one should consider for a rigorous definition of the inviscid limit. Namely, it is necessary to introduce a stochastic component in the regularization procedure. This can be performed by assuming, e.g., a vanishing small-scale noise or a small random perturbation of the viscosity~\cite{lorenz1969predictability,leith1972predictability,ruelle1979microscopic,eyink1996turbulence,mailybaev2015b}. 

The role of blowup configuration for the onset of spontaneous stochasticity can be seen within a framework of the classical theory of hyperbolic conservation laws. The classical scenario of shock formation requires an extension of the functional space in order to account for discontinuous solutions after the blowup. Our examples show that the nonlocal flux term may cause the instability in the process of shock formation, such that the physically relevant solution requires further extension of the functional space to stochastic distributions.
A different interpretation may be given, if one notices that the blowup is not a necessary ingredient of the theory, but rather one of the ways to create initial conditions that contain ``enough'' energy at all scales. For a shell model example, the theory directly extends to self-similar initial conditions $u_n \propto k_n^y$ for arbitrary $y > -1$. 

Initial conditions of that kind are relevant for the problem of developed turbulence, i.e., the inviscid limit of the incompressible Navier-Stokes equations. Finite-time blowup in the 3D incompressible Euler equations remains an open problem~\cite{gibbon2008three,gibbon2008three2,hou2008blowup}: numerical simulations suggest the blowup at a physical boundary~\cite{luo2013potentially}, while nearly exponential vorticity growth is typical for generic initial conditions with periodic boundary conditions~\cite{brachet1992numerical,agafontsev2015}. The latter means that the viscous range gets excited within logarithmic times $\propto \log \mathrm{Re}$ with respect to the Reynolds number, which makes the inviscid formulation with rough (weak) velocity fields physically relevant~\cite{Eyink2015}. The renormalization procedure proposed in this paper may be useful for understanding the spontaneous development of stochasticity from such turbulent initial conditions.  

\section*{Acknowledgments} 
This work was supported by the 
CNPq (grant 302351/2015-9) and the Program FAPERJ Pensa Rio (grant E-26/210.874/2014). The author is grateful to Gregory L. Eyink, Theodore D. Drivas and anonymous referees for useful discussions and suggestions.

\bibliographystyle{plain}
\bibliography{refs}
 
\end{document}